\documentstyle[aps,twocolumn,epsf,psfig]{revtex}

\begin{document}
\draft

\title {Effects of dimensionality and anisotropy on the Holstein polaron }

\author{Aldo~H.~Romero${^{0,1,2}}$,
David W. Brown${^3}$,
and Katja Lindenberg${^4}$}

\address
{${^1}$
Department of Chemistry and Biochemistry,\\
University of California, San Diego, La Jolla, CA 92093-0340} 

\address
{${^2}$
Department of Physics,\\
University of California, San Diego, La Jolla, CA 92093-0354}

\address
{${^3}$
Institute for Nonlinear Science,\\
University of California, San Diego, La Jolla, CA 92093-0402}

\date{\today} 

\maketitle

\begin{abstract}

We apply weak-coupling perturbation theory and strong-coupling perturbation theory to the Holstein molecular crystal model in order to elucidate the effects of anisotropy on polaron properties in $D$ dimensions.
The ground state energy is considered as a primary criterion through which to study the effects of anisotropy on the self-trapping transition, the self-trapping line associated with this transition, and the adiabatic critical point.
The effects of dimensionality and anisotropy on electron-phonon correlations and polaronic mass enhancement are studied, with particular attention given to the polaron radius and the characteristics of quasi-1D and quasi-2D structures.
Perturbative results are confirmed by selected comparisons with variational calculations and quantum Monte Carlo data.

\end{abstract}


\narrowtext

\footnotetext{Present address: Max-Planck Institut f\"{u}r Fest\-k\"{o}rper\-forschung, Heisenbergstr. 1, 70569 Stuttgart, Germany}

\section{Introduction}
\label{sec:intro}

Polarons are ubiquitous quasiparticles in deformable materials embodying the renormalizing effects of deformation quanta (phonons) on free carriers.
The effects that can appear depend on the strength of the electron-phonon coupling and on the relative time scales of the free electron motion and the relevant host vibrations, this latter relationship being subsumed in the common notion of adiabaticity.
Loosely speaking, at fixed adiabaticity weakly-coupled polarons may be spread over many lattice sites (``large'' or ``free'') while strongly-coupled polarons may be highly localized, even essentially completely collapsed (``small'' or ``self-trapped'').
Similarly, at fixed electron-phonon coupling strength very adiabatic polarons may be quite broad, while non-adiabatic polarons may be quite compact.
Part of the looseness in this characterization has to do with the nature of the self-trapping transition, and with dependences on the effective dimensionality of the host system.

Well-known and widely-invoked results tied to the adiabatic approximation
\cite{Rashba57a,Rashba57b,Holstein59a,Derrick62,Emin73,Sumi73,Emin76,Toyozawa80a,Schuttler86,Ueta86,Kabanov93,Silinsh94,Song96,Holstein81,Holstein88a,Holstein88b}
suggest that polarons in 1D should be qualitatively distinct from those found in 2D and 3D, and that even the notion of self-trapping should take on different meaning in low and high dimensions \cite{Rashba57a,Rashba57b,Holstein59a,Derrick62,Emin73,Sumi73,Emin76,Toyozawa80a,Schuttler86,Ueta86,Kabanov93,Silinsh94,Song96,Holstein81,Holstein88a,Holstein88b}.
The root of this lies in stability arguments suggesting that in 1D all polaron states should be characterized by finite widths, while in 2D and 3D polaron states may have either infinite radii (``free'' states at weak coupling) or finite radii (``self-trapped'' states at strong coupling).
The self-trapping transition is thus taken to mean the abrupt transition from delocalized ``free'' states characterized by the free electron mass to highly-localized ``self-trapped'' states characterized by strongly-enhanced effective masses.

The overall conclusion of this paper, on the other hand, consistent with a growing body of independent work \cite{DeRaedt83,DeRaedt84,Lagendijk85,Jeckelmann98a,Kornilovitch98a,Kornilovitch99}, is that the properties of higher-dimensional polarons are more qualitatively similar in most respects to those of 1D polarons than they are different, and that those distinctions that can meaningfully be drawn are only distantly related to the more familiar expectations outlined above.

Central in this subject is the notion of anisotropy, which figures particularly strongly in quasi-1D systems such as conducting polymers or in quasi-2D systems such as high-$T_c$ materials.
The nature of self-trapping in a quasi-1D system poses particularly potent challenges, since depending on one's stance one may reach divergent conclusions:  the polaron may self-trap, or it may not; it may be sharply localized, soliton-like, or free; its mass may be the free electron mass, may be weakly renormalized, or may be enhanced by orders of magnitude.
The resolution of such ambiguities lies not uniquely at the interface between 1D and 2D systems, but in a general understanding of the role of anisotropy in polaron structure, within which the clarification of the nature of self-trapping in quasi-1D systems is a byproduct.

An essential preliminary observation is that the self-trapping transition is not, in fact, an abrupt phenomenon in any dimension except in the adiabatic limit; at finite parameter values the physically-meaningful transition is more in keeping with a smooth, if rapid, ``crossover'' from polaron structures characteristic of the weak-coupling regime to structures characteristic of the strong-coupling regime.
The ``self-trapping line'' describing this transition can be located by criteria sensitive to changes in polaron structure; these may involve physical observables such as the polaron ground state energy and effective mass, or may rely upon more formal properties less accessible to direct physical measurement.
Here, we consider several physical observables at finite parameters as well as in asymptotic regimes.
Through these, we are able to characterize self-trapping in one, two, and three dimensions for any degree of anisotropy.

For the explicit calculations to follow, we use the Holstein Hamiltonian \cite{Holstein59a,Holstein59b} on a $D$-dimensional Euclidean lattice
\begin{eqnarray}
\hat{H} &=& \hat{H}_{kin} + \hat{H}_{ph} + \hat{H}_{int} ~, \\
\hat{H}_{kin} &=& - \sum_{ \vec{n} } \sum_{i=1}^D
J_i a_{\vec{n}}^{\dagger}
( a_{\vec{n}+\vec{\epsilon}_i} + a_{\vec{n}-\vec{\epsilon}_i} ) ~, \\
\hat{H}_{ph} &=& \hbar \omega \sum_{\vec{n}} b_{\vec{n}}^{\dagger} b_{\vec{n}} ~, \\
\hat{H}_{int} &=& - g \hbar \omega \sum_{\vec{n}} a_{\vec{n}}^{\dagger}
a_{\vec{n}} ( b_{\vec{n}}^{\dagger} + b_{\vec{n}} ) ~,
\end{eqnarray}
in which $a_{\vec{n}}^\dagger$ creates a single electronic excitation in the rigid-lattice Wannier state at site ${\vec{n}}$, and $b_{\vec{n}}^\dagger$ creates a quantum of vibrational energy  in the Einstein oscillator at site ${\vec{n}}$.
All sums are understood to run over the entire infinite, periodic, $D$-dimensional lattice.
Because there is no phonon dispersion in this model, and because the electron-phonon coupling is strictly local, it is in $\hat{H}_{kin}$ where lattice dimensionality and structure have their greatest influence; the $J_i$ are the nearest-neighbor electronic transfer integrals along the primitive crystal axes, and the $\hat{\epsilon}_i$ are unit vectors associated with the primitive translations.
The above model encompasses all Bravais lattices, with the different lattice structures appearing only in the relative values of the hopping integrals $J_i$.
For simplicity in the following, we use terms appropriate to orthorhombic lattices in which conventionally $i = x$, $y$, or $z$; however, all results hold for lattices of lower symmetry with appropriate transcription of these labels to those of the primitive axes.

For some purposes in this paper, we qualify and quantify anisotropy through a vector
\begin{equation}
\vec{J} = J_x \hat{\epsilon}_x + J_y \hat{\epsilon}_y + J_z \hat{\epsilon}_z
\end{equation}
whose orientation in a Cartesian system can be used to objectively quantify anisotropy.
In these terms, an isotropic property is one depending only on the modulus $ | \vec{J} | = (\vec{J} \cdot \vec{J})^{1/2}$.

For other purposes, however, it is convenient to think of dimensions being turned ``on'' or ``off'' according to whether particular $J_i$ are finite or vanishing.
Anisotropy can then be tuned by varying selected $J_i$ in the interval $(0,J]$.
In several illustrations to follow, we do this sequentially, so that dimensions are ``turned on'' one by one, arriving ultimately at the isotropic $D$-dimensional case in which $J_i = J$ along all axes.
In the following, we reserve the unsubscripted scalar symbol $J$ to represent common magnitude of all $J_i$ in an isotropic case ($J = J_i = | \vec{J} | / \sqrt{D}$).
This manner of tuning dimensionality does not isolate anisotropy, however, since changing one $J_i$ keeping others fixed changes both the orientation and modulus of $\vec{J}$.

From either perspective, it is such tuning between dimensions by continuously varying the anisotropy that is the physically-meaningful concept in most situations characterized as quasi-1D ($J_x >> J_y ,~ J_z$) or quasi-2D ($J_x,~ J_y >> J_z$).

Quite apart from such very direct quantifications of anisotropy, another quantity that arises naturally in the following is the sum of the transfer integrals along each axis
\begin{equation}
{\cal{J}} \equiv \sum_{i=1}^D J_i ~~\propto~~ Tr ~ {\bf M_0^{-1} } ~,
\end{equation}
in which ${\bf M_0^{-1} }$ is the reciprocal effective mass tensor of the free electron (see Eq. \ref{eq:masstensor} {\it ff.}); in the isotropic case, ${\cal{J}} = D J$.

Using weak-coupling perturbation theory (WCPT) identifying the unperturbed Hamiltonian as $\hat{H}_0 = \hat{H}_{kin} + \hat{H}_{ph}$ and the perturbation as $\hat{H}' = \hat{H}_{int}$ \cite{Alexandrov95,Nakajima80,Mahan93,Romero98g}, one can show that the form of the polaron energy band at weak coupling in $D$ dimensions is given by
\begin{equation}
E(\vec{\kappa}) = E_{WC}^{(0)} ( \vec{\kappa} ) + E_{WC}^{(2)} (\vec{\kappa}) + O\{g^4 \} ~,
\label{eq:wcpt}
\end{equation}
where
\begin{eqnarray}
E_{WC}^{(0)} ( \vec{\kappa} ) &=& - \sum_{i=1}^{D} 2J_i \cos \kappa_i  ~, 
\label{eq:wcpt0} \\
E_{WC}^{(2)} ( \vec{\kappa} ) &=& \nonumber \\
- g^2 && \!\!\!\!\!\!\!\! \hbar^2 \omega^2 \!\! \int_0^{\infty} \!\! \!\! \!\! dt ~ e^{ - \hbar \omega t} \prod_{i=1}^D  e^{-2J_i t \cos \kappa_i} I_0 (2J_i t) ,
\label{eq:wcpt2}
\end{eqnarray}
in which $I_n (z)$ is the modified Bessel function of order $n$.

Using strong-coupling perturbation theory (SCPT) following the Lang-Firsov transformation, identifying the unperturbed Hamiltonian as $\tilde{H}_0 = \tilde{H}_{ph} + \tilde{H}_{int}$ and the perturbation as $\tilde{H}' = \tilde{H}_{kin}$) \cite{Alexandrov95,Romero98g,Marsiglio95,Stephan96}, one finds
\begin{equation}
E(\vec{\kappa}) = E_{SC}^{(0)} ( \vec{\kappa} ) + E_{SC}^{(1)} (\vec{\kappa}) + E_{SC}^{(2)} (\vec{\kappa}) + O \left\{ \frac {\tilde{J}^3} {\hbar^3\omega^3} \right\} ,
\label{eq:scpt}
\end{equation}
where $\tilde{J}$ is an effective, {\it dressed} tunneling parameter that may be either comparable to or much smaller than the bare $J$ depending on regime, and
\begin{eqnarray}
E_{SC}^{(0)} ( \vec{\kappa} ) &=& - g^2
\label{eq:scpt0} \\
E_{SC}^{(1)} ( \vec{\kappa} ) &=& - e^{-g^2} \sum_{i=1}^D 2J_i \cos \kappa_i
\label{eq:scpt1} \\
E_{SC}^{(2)} ( \vec{\kappa} ) &=& - e^{ -2 g^2} f(2 g^2) \sum_{i=1}^D 2 J_i^2 \nonumber \\
&& - e^{ -2 g^2} f(g^2) \sum_{i=1}^D 2 J_i^2 \cos 2\kappa_i \nonumber \\
&& - e^{ -2 g^2} f(g^2) \sum_{i \ne j}^D J_i J_j \cos \kappa_i \cos \kappa_j
\label{eq:scpt2} \\
f(y) &=& {\rm{Ei}}(y) - \gamma - \ln(y)
\label{eq:fofy}
\end{eqnarray}
where $\gamma$ is the Euler's constant and ${\rm{Ei}}(y)$ is the exponential
integral.

\begin{figure}[htb]
\begin{center}
\leavevmode
\epsfxsize = 3.2in
\epsffile{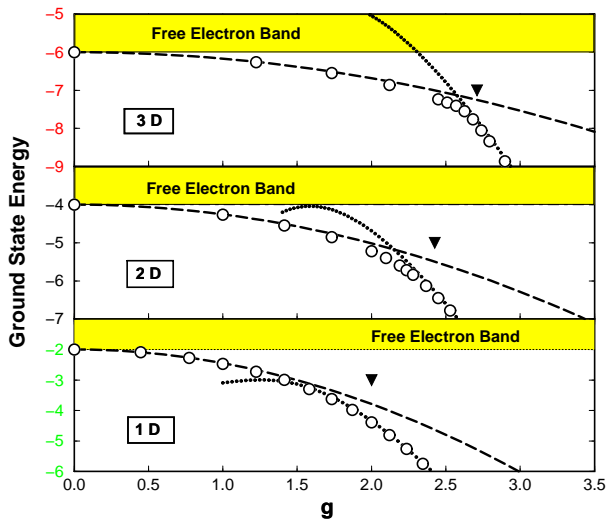}
\end{center}
\caption{
Ground state energies (in units of $\hbar \omega$) in 1D, 2D, and 3D for $J/\hbar\omega = 1$.
Solid line:  Strong-coupling perturbation theory through second order.
Dashed line:  Weak-coupling perturbation theory through second order.
Circles:  Quantum Monte Carl data kindly provided by P. E. Kornilovitch \protect \cite{Kornilovitch98a,Kornilovitch99} \protect.
Triangles:  Estimated self-trapping points $g_{ST}$ as discussed in Section \ref{sec:finite}.
}
\label{fig:grnd123d}
\end{figure}

In Figure~\ref{fig:grnd123d}, we show results for the ground state energy in 1D, 2D, and 3D for weak-coupling perturbation theory through second order and strong-coupling perturbation theory through second order, together with corresponding results of quantum Monte Carlo simulation \cite{Kornilovitch98a,Kornilovitch99}.
This comparison shows that in any dimension weak-coupling and strong-coupling perturbation theory are both quite good up to a relatively small interval around the ``knee'' that is associated with the self-trapping transition.
The real knee as discernible in the quantum Monte Carlo data characteristically falls to the strong-coupling side of the intersection (or near-intersection) of the WCPT and SCPT curves, but to the weak-coupling side of $g_{ST} $ as discussed in Section \ref{sec:finite}
\cite{Romero98c}.
(These systematic offsets are symptomatic of the smooth nature of the physically-meaningful transition, as discussed in the next section.)

Beyond validating both weak and strong-coupling perturbation theory as used in this paper, Figure~\ref{fig:grnd123d} also holds a message regarding the qualitative character of self-trapping in different dimensions.
That message is that the occurrence of self-trapping is qualitatively similar in one, two, and three dimensions, with the primary qualitative changes being that the transition systematically increases in abruptness and shifts to stronger coupling as dimensionality is increased.

\section{Self-trapping preliminaries}
\label{sec:prelim}

A central result upon which the following sections build is the concept of a self-trapping line as contained in the empirical curve
\begin{equation}
g_{ST} [1] = 1 + \sqrt{J/\hbar\omega}
\label{eq:gst}
\end{equation}
that has been found to accurately characterize the transition between the small and large polaron regimes in one dimension.
This curve was inferred through the application of objective criteria to physical properties such as the polaron effective mass \cite{Romero98e}, ground state energy, kinetic energy, phonon energy, electron-phonon interaction energy \cite{Romero98c}, and electron-phonon correlation function \cite{Romero99b}.
Although this simple construct consistently describes a wealth of data drawn from multiple polaron properties obtained by our own and independent methods, it is well to stress what the above relation is {\it not}, since the same limitations apply to other constructs we are led to in the balance of this paper.

The self-trapping curves we address do not describe a phase transition, nor even the exact location of the objectively-determined point of crossover implicit in any one physical property.
Different physical properties generally signal the occurrence of self-trapping at distinct, though systematically and tightly clustered points on the polaron phase diagram.
The empirical self-trapping line is not intended to describe any one property exactly, but to accurately describe the central trend of clusters of transition properties over a large range of the polaron parameter space.
Self-trapping loci drawn from observations of different physical properties thus track the empirical trend line with their own systematic deviations that narrow as the adiabatic limit is approached.

Clearly, the form of $g_{ST} [1]$ has not been derived from first principles.
Indeed, one can easily be persuaded from approximate descriptions of the problem that the initial dependence of $g_{ST}[1]$ upon $J/\hbar\omega$ is most likely not singular, but regular, if perhaps steep \cite{Brown99}.
Our retention of the square root in $g_{ST} [1]$ and its higher-dimensional generalizations developed below thus reflects not an assertion of singular physical behavior, but merely an economy of phenomenology.

\section{Self-Trapping Transition in the Scaling Regime}
\label{sec:scaling}

The location of the self-trapping line can be estimated under the practical assumption that the self-trapping transition should lie near the crossover from the weak-coupling regime to the strong-coupling regime; here, specifically, by the intersection of the ground state energy curves as given by the leading orders of each perturbation theory

This is an imperfect assumption, since the errors in both weak-coupling perturbation theory and strong-coupling perturbation theory increase in absolute terms as the transition is approached.
Absolute precision in the perturbative energy is not required, however, in order to accurately locate the transition in parameter space.
Handled carefully, we {\it can} expect a WCPT/SCPT crossing condition to capture dependences that are asymptotically correct in the adiabatic strong-coupling limit, provided that relative errors in the appropriate quantities remain controlled.
Since we are limited to the low orders of perturbation theory, we necessarily depend upon there being no unexpected surprises lurking in the higher orders of either weak or strong-coupling perturbation theory that upset the scaling relationships evident in the leading orders; that such might, in principle, occur is a caveat, however unlikely, that must attach to our arguments.

To this end, we consider the adiabatic strong-coupling limit where the self-trapping line coincides with the adiabatic critical point; i.e., where the smooth physical transition steepens critically.
The composite parameter
\begin{equation}
\lambda = \frac {E_{SC}^0 (0)} {E_{WC}^0 (0)} = \frac {g^2 \hbar \omega} {2 {\cal{J}} }
\label{eq:lambdadef}
\end{equation}
appears frequently in discussions of this regime because it embodies the essential scaling relationship characterizing the adiabatic strong-coupling limit.
This dominant scaling relationship, $g^2 \sim {\cal{J}}/\hbar\omega$, guides our application of perturbation theory to the estimation of the location of the self-trapping transition.
We note that $\lambda$ depends not on the modulus of $\vec{J}$ but on ${\cal{J}}$, the sum of the components $J_i$, and thus by its very definition $\lambda$ includes a dependence on anisotropy.

\begin{figure}[htb]
\begin{center}
\leavevmode
\epsfxsize = 3.2in
\epsffile{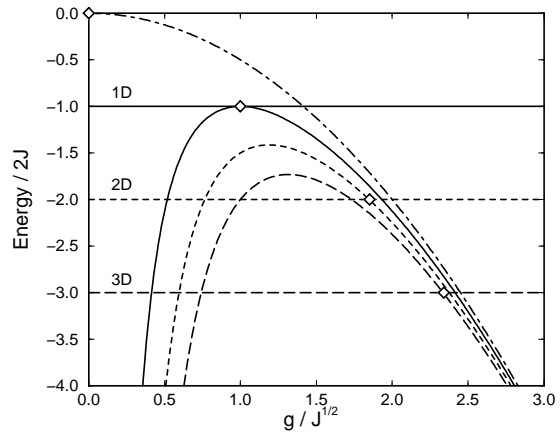}
\end{center}
\caption{
Ground state energies in the adiabatic strong\-coupling limit for the isotropic case.
Horizontal lines indicate the l.h.s. of (\ref{eq:adiabaticx}) and curved lines indicate the r.h.s. of (\ref{eq:adiabaticx}).
Solid: 1D.
Short-dashed: 2D.
Long-dashed: 3D.
Symbols indicate $\lambda_c $.
The chain-dotted curve ($-g^2$) can be viewed as corresponding to the $D=0$ case.
}
\label{fig:critical}
\end{figure}

The perturbative results (\ref{eq:wcpt}) - (\ref{eq:fofy}) can be used to infer the expected value of the adiabatic critical point by retaining only those terms that dominate in the adiabatic strong-coupling regime; these are terms of comparable, leading magnitude when both $g$ and ${\cal{J}}/\hbar\omega$ are large such that $g^2 \sim {\cal{J}}/\hbar\omega$.
The weak-coupling correction $E_{WC}^{(2)}$ is $O\{ - \lambda \}$ at large ${\cal{J}}$ and is thus negligible relative to $E_{WC}^{(0)}$, which is $-2{\cal{J}}$.
The strong-coupling correction $E_{SC}^{(1)}$ is exponentially small (in $g^2$) relative to $E_{SC}^{(0)}$ and is thus negligible in the adiabatic strong-coupling regime.
Of the several contributions to $E_{SC}^{(2)}$ appearing in (\ref{eq:scpt2}), only the term containing $f(2g^2)$ is not exponentially small, and of the terms contributing to this non-exponential contribution, only a single, dominant term remains non-vanishing in the adiabatic strong-coupling regime.
Thus, combining all non-vanishing terms through second order of both weak- and strong-coupling perturbation theory, the crossing condition that obtains is
\begin{equation}
-2{\cal{J}} = -g^2 \hbar\omega - \frac {| \vec{J} | ^2} {g^2 \hbar\omega} ~.
\label{eq:adiabaticx}
\end{equation}
The graphical solution of (\ref{eq:adiabaticx}) is indicated in Figure~\ref{fig:critical}.

In obtaining (\ref{eq:adiabaticx}), we are using perturbative results in extreme limits that may not obviously lie within the scope of the retained orders of either perturbation theory, or in principle may even lie beyond the scope of one or the other perturbation theory taken to all orders.
Although the legitimacy of our arguments in this regard is beyond the scope of any available proof, it is not unsupported; that the trends in the true ground state energy are consistent with the scaling properties used to obtain (\ref{eq:adiabaticx}) is evident in the results of multiple independent non-perturbative methods on both sides of the self-trapping transition \cite{Romero98a}.
Such studies are necessarily at finite parameter values, however, and though confirmatory cannot in themselves prove that these trends continue unabated into the adiabatic limit.
In Appendix A, we provide a discussion of WCPT in particular, showing how it is feasible that WCPT may continue to be valid in the scaling regime despite what may appear to be essentially strong coupling.

We note that it is the second-order SCPT contribution on the r.h.s. of (\ref{eq:adiabaticx}) that is the crucial element in much of the discussion that follows.
If one fails to capture this contribution to the crossing condition, the self-trapping criterion that results is simply $\lambda = 1$.
This is, in fact, a widely-asserted self-trapping condition and is not grossly incorrect in many cases; however, considerable structure is lost to the casualness with which this estimate is often used, and the potential exists for significant quantitative errors if applied to the inappropriate regimes.

In terms of the composite parameter $\lambda$ and according to the {\it full} condition (\ref{eq:adiabaticx}), the adiabatic critical point in any dimension is given by
\begin{equation}
\lambda_c = \frac 1 2 \left[ 1 \pm \sqrt{ 1-  | \vec{J} |^2 /{\cal{J}}^2} ~ \right] ~.
\label{eq:lambdacd}
\end{equation}
Of the two roots, it is the larger, $(+)$ root that is the physically meaningful one, yielding the {\it isotropic} (superscripts ``$i$'') critical values
\begin{equation}
\lambda_c^i [D] = \frac 1 2 \left[ 1 + \sqrt{1-D^{-1}} \right] ~,
\end{equation}
\begin{equation}
\lambda_c [1] = 0.5 ~,~~ \lambda_c^i [2] = 0.8536... ~,~~ \lambda_c^i [3] = 0.9082... ~,
\end{equation}
\begin{equation}
\lambda_c [0] = 0 ~,~~ \lambda_c^i [ \infty ] = 1.0 ~.
\end{equation}
The $(-)$ root, besides implying an unmeaningful dependence of the ground state energy on parameters, would imply a $\lambda_c $ that {\it decreases} with increasing dimensionality, contrary to considerable evidence, including the quantum Monte Carlo data shown in Figures~\ref{fig:grnd123d} and \ref{fig:mass123d}.

\begin{figure}[htb]
\begin{center}
\leavevmode
\epsfxsize = 3.2in
\epsffile{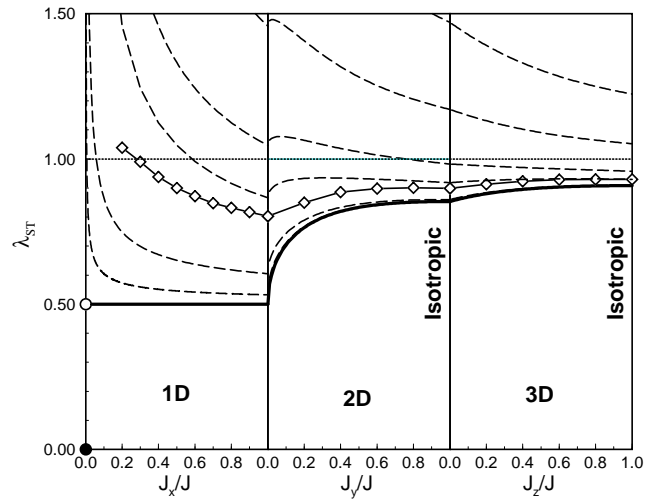}
\end{center}
\caption{
Dependence of $\lambda_{ST} $ on anisotropy, and related curves.
{\bf Solid line:}
$\lambda_{c} $ (adiabatic limit) as anisotropy is tuned.
Bullet at lower left:  $\lambda_{c} [0] =0$.
Left Panel:
$\lambda_c$ such that $J_y = J_z = 0$, increasing $J_x$ from $0$ to $J$.
Center Panel:
$\lambda_c$ such that $J_x = J, ~J_z = 0$, increasing $J_y$ from $0$ to $J$.
Right Panel:
$\lambda_c$ such that $J_x = J_y = J$, increasing $J_z$ from $0$ to $J$.
{\bf Connected diamonds:}
Solution of the full weak/strong criterion (\ref{eq:crossing}) for $J/\hbar\omega = 10$.
{\bf Dashed lines:}
The estimated $\lambda_{ST}$ as given by (\ref{eq:gst}) for $J/\hbar\omega = 1,~ 2,~ 5,~ 10,~ 100,~1000$, from top to bottom.
}
\label{fig:lambdac123}
\end{figure}

\begin{figure}[htb]
\begin{center}
\leavevmode
\epsfxsize = 3.2in
\epsffile{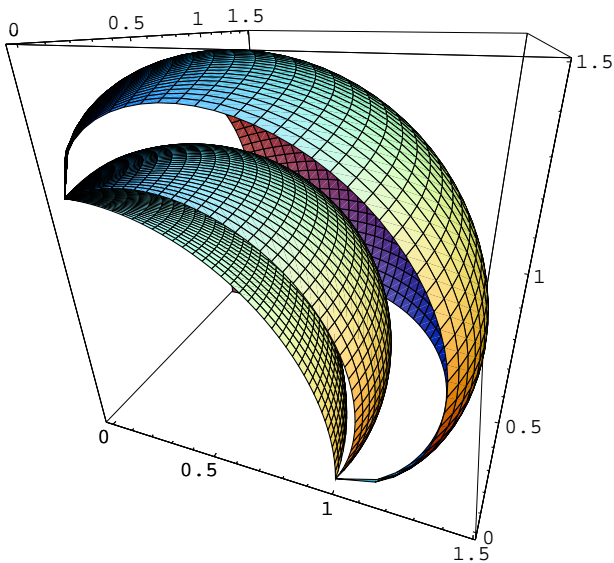}
\end{center}
\caption{
Dependence of self-trapping on anisotropy.
The Cartesian axes are the positive half-axes of the $\{ J_x , J_y , J_z \}$ system.
The innermost surface is an octant of the unit sphere exhibited as a reference surface reflecting isotropic dependence on $| \vec{J} |$ only.
The outermost surface is defined such that the radial distance from the origin represents the ratio $g_{c} /g_{c} [1]$ as in (\ref{eq:gratio}).
The interleaved surface is defined such that the radial distance from the origin reflects the anisotropy implicit in the condition $\lambda_c = const.$ at fixed $| \vec{J} |$.
}
\label{fig:gratio}
\end{figure}

The dependence of the adiabatic critical point on anisotropy contains interesting structure (see Figure~\ref{fig:lambdac123}):
In all of 1D and in each weakly anisotropic case for $D > 1$, the dependence of $\lambda_c $ (solid line) on the anisotropy is essentially flat; thus, in the generic case of ordinary bulk materials with only modest anisotropies, $\lambda_c $ would appear to change significantly with dimensionality but to be essentially insensitive to underlying anisotropy.
On the other hand, in the case of ``low-dimensional'' materials characterized by weak tunneling into one or more transverse dimensions (e.g., quasi-1D scenarios with one or two transverse dimensions, or quasi-2D scenarios with one), the weakly-involved dimensions have relatively strong effects on $\lambda_c $:
The transition between zero dimensions and any higher dimensional case is marked by a jump discontinuity in the dependence of $\lambda_c$ on any $J_i$.
The transition between one dimension and any higher-dimensional case is marked by a square-root singularity in dependence of $\lambda_c$ on the transverse $J_i$.
The transition between successive higher dimensional cases is generically smooth, however, with the appearance of any singularity being dependent on the manner in which dimensionality is tuned (see Figure~\ref{fig:gratio} {\it ff.}).

The abscissa in Figure~\ref{fig:lambdac123} is essentially the quantity ${\cal{J}}/J$ as ${\cal{J}}$ ranges from $0$ to $3J$ according to the particular scheme chosen for sequentially ``turning on'' higher dimensions.
It is clear that while the casual criterion $\lambda \sim 1$ constitutes a fair order-of-magnitude characterization of the occurrence of self-trapping in bulk materials in the adiabatic limit, there is considerable qualitative structure missed.

The results shown in Figure~\ref{fig:lambdac123}, though quite general in character, depend on the particular manner in which the parameters $\{ J_x , J_y , J_z \}$ are varied relative to each other; in particular, we note that the manner in which these are varied in Figure~\ref{fig:lambdac123} does not isolate the anisotropy.
We can obtain a more global view of self-trapping in higher dimensions while simultaneously isolating the anisotropy dependence by considering not the composite parameter $\lambda_c $, but the more elementary coupling parameter $g_c $ contained within it according to (\ref{eq:lambdadef}).
That is,
\begin{equation}
g_c = \sqrt{ 2 {\cal{J}} \lambda_c / \hbar \omega} ~,
\label{eq:criticalgst}
\end{equation}
This critical value of the coupling constant in the adiabatic limit depends on both the intensity of tunneling $| \vec{J} |$ and on the anisotropy.
The dependence on the anisotropy can be isolated, however, in the normalized quantity
\begin{equation}
\frac {g_c } {g_c [1]} = \left\{ \frac { {\cal{J}} } {| \vec{J} | } \left[ 1 + \sqrt{ 1- | \vec{J} |^2 /{\cal{J}}^2} ~ \right] \right\} ^{1/2}
\label{eq:gratio}
\end{equation}
in which $g_c [1] = \sqrt{| \vec{J} /\hbar\omega  |}$ represents the critical coupling parameter in the one-dimensional case subject to the condition that the 1D tunneling parameter is fixed at the value $| \vec{J} |$ appropriate to the $D$-dimensional case.
While dependent on each of $J_x$, $J_y$, and $J_z$, the ratio (\ref{eq:gratio}) is independent of $| \vec{J} |$ and depends only on the angular variables in a spherical polar coordinate representation of the $\{J_x , J_y , J_z \}$ system.
Eq. \ref{eq:gratio} thus describes a surface having the interpretation that the radial distance from the origin is the factor by which the critical coupling constant $g_c $ in $D$ dimensions exceeds the critical value in one-dimension ($g_c [1]$) having the same intensity of tunneling.
This surface is plotted in Figure~\ref{fig:gratio}, together with an isotropic (spherical) reference surface, and a surface corresponding to the condition $\lambda_c = const$.
This latter surface shows that the oft cited condition $\lambda_c \sim 1$ contains implicit anisotropy; however, it is evident that the real anisotropy of self-trapping is even greater than might be inferred from this common rule of thumb.

The presentation of $\lambda_c $ shown in Figure~\ref{fig:lambdac123} corresponds to a particular transit of the $g_c $ surface seen in Figure~\ref{fig:gratio}:
The 0D case can be considered to occupy the origin, and the 1D cases correspond to the three corners of the displayed surface.
The ``turning on'' of the second dimension according to the scheme of Figure~\ref{fig:lambdac123} corresponds to movement along the edge of the displayed surface to the midpoint of that edge corresponding to the 2D isotropic case.
The subsequent ``turning on'' of the third dimension corresponds to movement
perpendicular from this edge along a straight line (geodesic) to the center of the surface corresponding to the 3D isotropic case.
This comparison shows in particular (as may be proven analytically):
1) that both the jump discontinuity between 0D and higher dimensions and the square-root singularity between 1D and higher dimensions are generic features, not dependent on the manner or sequence with which transverse dimensions are ``turned on'', and
2) that the less-singular feature seen in Figure~\ref{fig:lambdac123} at the transition from 2D to 3D is {\it not} generic, but appears only because dimensions in Figure~\ref{fig:lambdac123} were turned on {\it sequentially}.

Thus, for a given $| \vec{J} |$, we can distinguish three regimes based on sensitivity to anisotropy:

$g > g_c^i$;
there are no large polaron states at any degree of anisotropy.

$g < g_c [1]$;
there are no small polaron states for any degree of anisotropy.

$g_c [1] < g < g_c^i$;
large polaron states exist for sufficiently isotropic tunneling, small polaron states exist for sufficiently anisotropic tunneling, and these regimes are separated by a self-trapping transition as a function of anisotropy at fixed $| \vec{J} |$ and $g$.
This effect of self-trapping as a function of anisotropy alone can be understood in terms of the size, shape, and content of the phonon cloud.
As discussed in Appendix A, the more isotropic and higher-dimensional polaron scenarios are characterized by phonon clouds that are spread over the largest volumes of space and contain the fewest numbers of phonons.
With increasing anisotropy, the polaron cloud grows more compressed, occupying smaller volumes of space, but being occupied by larger numbers of phonons.
If this anisotropy-driven compression can proceed sufficiently far, the number of phonons in the phonon cloud can be driven sufficiently high that self-trapping can occur.

\section{Self-Trapping away from the Adiabatic Strong-Coupling Limit}
\label{sec:finite}

At general parameter values away from extreme limits, accurate estimations of the location of the polaron self-trapping line are scarce.
Until rather recently, estimates even in one dimension were largely casual rules of thumb.
As noted above, a frequently-encountered characterization holds that self-trapping occurs when $\lambda \sim 1$; this condition is often supplemented by the condition $g > 1$ acknowledging that the strong-coupling theory from which the $\lambda$ condition arises is not expected to hold to arbitrarily weak coupling.

\begin{figure}[htb]
\begin{center}
\leavevmode
\epsfxsize = 3.2in
\epsffile{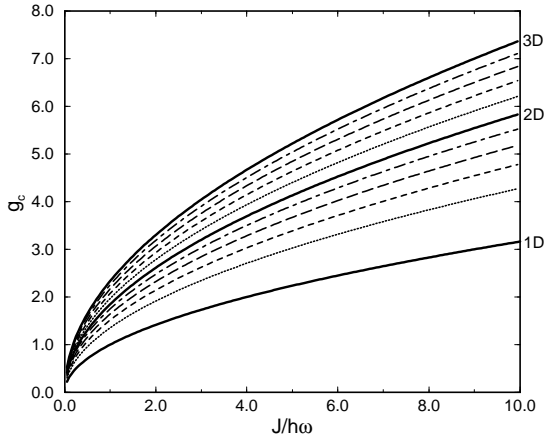}
\end{center}
\caption{
Dependence of the self-trapping line on anisotropy under the assumption
$\lambda_{ST} = \lambda_c $.
The bold solid lines correspond to the isotropic cases in (from bottom to top) 1D, 2D, and 3D.
The textured lines between the 1D and 2D isotropic cases correspond to (from bottom to top) $J_x = J$, $J_z = 0$, with $J_y = 0.2J, ~0.4J, ~0.6J$, and $0.8J$.
The textured lines between the 2D and 3D isotropic cases correspond to (from bottom to top) $J_x = J_y = J$, with $J_z = 0.2J, ~0.4J, ~0.6J$, and $0.8J$.
}
\label{fig:lsteqlc}
\end{figure}

We can improve on the common rule of thumb by identifying the self-trapping transition not with a single fixed value of $\lambda$ (e.g., unity) but with the critical value obtaining in the adiabatic limit for the particular dimension and $\vec{J}$ appropriate to each unique circumstance (i.e., $\lambda \sim \lambda_c$).
In so doing, we capture all the structure evident in $\lambda_c $ (Figures \ref{fig:lambdac123} and \ref{fig:gratio}) and make the preliminary assumption that the scaling relationships that characterize the adiabatic limit hold to a meaningful degree at moderate parameter values; i.e, we may consider extrapolation of critical scaling relationships to finite parameter values.
The implications of such an assumption for the elementary coupling parameter
$g_c $
are shown in Figure~\ref{fig:lsteqlc}.
The shifting of these estimated self-trapping lines with anisotropy is a direct reflection of the anisotropy of $\lambda_c $.
The shifting of these estimated self-trapping lines with anisotropy is a direct reflection of the anisotropy of $\lambda_c$; it is this qualitative character of the mutual relationships among self-trapping curves of differing anisotropies that we expect to be largely preserved as necessary corrections are made.
The need for further correction is evident, for example, in that the 1D example in Figure~\ref{fig:lsteqlc} differs substantially in absolute terms from the 1D empirical curve (\ref{eq:gst}) although the two are qualitatively quite similar.
Moreover, all of the curves displayed in Figure~\ref{fig:lsteqlc} violate the ancillary condition $g > 1$ at small ${\cal{J}}/\hbar\omega$, reflecting the expected eventual failure of extrapolation from the adiabatic strong-coupling regime.

\begin{figure}[htb]
\begin{center}
\leavevmode
\epsfxsize = 3.2in
\epsffile{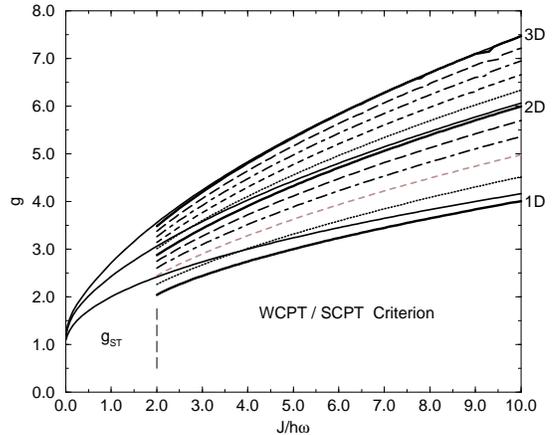}
\end{center}
\caption{
Truncated curves show the dependence of the self-trapping line on anisotropy using the complete weak/strong condition (\ref{eq:crossing}) as a self-trapping criterion; these curves are clipped below $J/\hbar \omega = 2$ because crossings of weak- and strong-coupling results begin to disappear (see the 1D case of Figure~\ref{fig:grnd123d}).
The untruncated solid curves are the empirical $g_{ST}$ of (\ref{eq:gstd}) in the isotropic cases of each dimensionality.
The bold solid lines correspond to the isotropic cases in (from bottom to top) 1D, 2D, and 3D.
The textured lines between the 1D and 2D isotropic cases correspond to (from bottom to top) $J_x = J$, $J_y = 0.2J, ~0.4J, ~0.6J$, and $0.8J$, and $J_z = 0$.
The textured lines between the 2D and 3D isotropic cases correspond to (from bottom to top) $J_x = J_y = J$, and $J_z = 0.2J, ~0.4J, ~0.6J$, and $0.8J$.
}
\label{fig:phase123}
\end{figure}

We should be able to improve on this estimate by using a more complete weak/strong condition (\ref{eq:crossing}) employing the complete results of both perturbation theories through second order as given in (\ref{eq:wcpt}) - (\ref{eq:fofy}), thus objectively capturing non-adiabatic corrections implicit in those terms that do not contribute in the adiabatic limit.
Thus we consider the condition
\begin{equation}
E_{WC}^{(0)} (0) + E_{WC}^{(2)} (0) = E_{SC}^{(0)} (0) + E_{SC}^{(1)} (0) + E_{SC}^{(2)} (0) ~.
\label{eq:crossing}
\end{equation}
This refinement yields estimated self-trapping lines as illustrated by the truncated curves in Figure~\ref{fig:phase123}; these curves are truncated (arbitrarily at $J/\hbar \omega =2$) because intersections of WCPT and SCPT begin to disappear at lower values of $J/\hbar \omega$, as can be seen in the 1D panel of Figure~\ref{fig:grnd123d}.

The effects of including non-adiabatic corrections depend on dimensionality, anisotropy, and ``distance'' from the adiabatic limit:
1)~The self-trapping curves describing 1D and quasi-1D cases shift strongly to stronger coupling values, suggesting a corrective shift of order unity at essentially all $\vec{J}$.
2)~The self-trapping curves describing 2D and 3D cases shift only weakly at moderate adiabaticity and more weakly with increasing adiabaticity.
3)~Except for strong corrections in the quasi-1D regime, the qualitative character of the dependence of self-trapping on anisotropy is little affected by non-adiabatic corrections.
4)~At low adiabaticity, all self-trapping curves shift to stronger coupling values in a manner and to a degree consistent with a condition $g > 1$ at $\vec{J} =0$ rather than the condition $g>0$ suggested by adiabatic scaling.

Gathering all the implications of the above together, we are led to extend our 1D empirical curve (\ref{eq:gst}) describing the one-dimensional self-trapping to the general case describing any dimension and any degree of anisotropy.
To do this we combine:
a) the empirical curve $g_{ST} [1]$ that effectually characterizes the one-dimensional case,
b) the adiabatic critical curve $g_c $ that effectually characterizes the higher-dimensional, higher-adiabaticity regime, and
c) the adiabatic critical parameter $\lambda_c $ that compactly describes the qualitatively distinct characteristics of the low and high dimensionalities.

From such considerations we are led to a family of empirical curves
\begin{equation}
g_{ST} \sim (1+{\cal{J}}/\hbar\omega)^{(\lambda_c [1] - \lambda_c ) \cdot ({\cal{J}} / | \vec{J} | )} + g_c
\label{eq:gstd}
\end{equation}
in which all quantities have been previously defined.
This family of curves is not derived from any theory, and, apart from the 1D case, is not backed by a large body of independent high-quality data since such data is quite sparse at the present time.
What high-quality data does exist at the present time is quantitatively consistent with this family of curves in the same fashion that an abundance of high-quality 1D data has been found consistent with $g_{ST} [1]$ (see Figures \ref{fig:grnd123d} and \ref{fig:mass123d}).
In keeping with the discussion of Section \ref{sec:prelim}, we have not attempted to regularize square root dependences that arise naturally in the adiabatic limit, but which are most likely softened with decreasing adiabaticity.
The utility of (\ref{eq:gstd}) lies in compactly and simply describing the apparent and mutually consistent trends in a large volume of results of independent methods and arguments, providing meaningful estimates for the location of the self-trapping transition in any dimension for any degree of anisotropy or adiabaticity.
This estimated $g_{ST} $ is compared with quantum Monte Carlo data for the ground state energy in Figure~\ref{fig:grnd123d} and effective mass in Figure~\ref{fig:mass123d}.

In Figure~\ref{fig:lambdac123}, we have included several curves (dashed lines) corresponding to $\lambda_{ST} \equiv g_{ST}^2/2{\cal{J}}$, using (\ref{eq:gstd}) for $J/\hbar \omega = 1, ~2, ~5, ~10, ~100, ~1000$.
These curves indicate how we expect the physically meaningful self-trapping line at finite parameters as estimated by (\ref{eq:gstd}) to be related to the results of the adiabatic limit.
Figure~\ref{fig:lambdac123} shows that the higher-dimensional, more isotropic cases converge toward their adiabatic limits more rapidly than do lower-dimensional, more anisotropic cases.
This convergence in one dimension is particularly poor, with significant deviations from the adiabatic limit persisting for $J/\hbar\omega > 1000$, by which point the higher-dimensional cases have converged beyond plotting precision.

Viewed collectively, the dashed curves of Figure~\ref{fig:lambdac123} also show that the composite parameter $\lambda$ does not provide a very natural or even qualitatively self-consistent characterization of the self-trapping transition over the whole of the adiabatic regime ($J/\hbar\omega > 1/4$).
In the far adiabatic regime, where we may take $\lambda_c$ to fairly characterize the location of the self-trapping transition (solid curve in Figure~\ref{fig:lambdac123}), one may be led to conclude that large polarons are relatively more stable in higher dimensions and at weaker anisotropies since the occurrence of self-trapping is found to shift to larger values of $\lambda$ in these regimes.
On the other hand, at more moderate degrees of adiabaticity (e.g., $J/\hbar\omega = 1, 2, 5$ in Figure~\ref{fig:lambdac123}) one is led by the same reasoning to conclude that large polarons are relatively {\it less} stable in higher dimensions and at weaker anisotropies since the occurrence of self-trapping is found to shift to {\it lower} values of $\lambda$ in these regimes.
In particular, one of the most actively-investigated cases in contemporary studies is the ``typical'' scenario with $J/\hbar\omega$ of order unity; Figure~\ref{fig:lambdac123} shows that in terms of $\lambda$, the self-trapping trends in this case are quite distinct from those found in the adiabatic limit, certainly complicating the interpretation of results.

From Figure~\ref{fig:phase123}, on the other hand, based on the more elementary coupling parameter $g$ appearing directly in the Hamiltonian, one is led to conclude that large polarons are everywhere relatively more stable in higher dimensions and at weaker anisotropies since the self-trapping line shifts to larger values of $g$ as these trends are followed regardless of the degree of adiabaticity.
These trends in $g$ are qualitatively similar and uniform for all degrees of anisotropy and adiabaticity, whereas the same trends in $\lambda$ vary strongly with regime.
For the same reasons that $\lambda$ is a convenient parameter with which to characterize polarons in the far adiabatic regime, it proves to be an inconvenient parameter in the broader context of the problem away from the adiabatic limit.

\section{Correlation Function and Polaron Radius}
\label{sec:corr}

In view of the local nature of the electron-phonon coupling in the Holstein model, the spatial extent of the polaron can be characterized quite directly through an analysis of electron-phonon correlations.
This can be done using a correlation function that has been long and widely used to characterize polaron size in $D$ dimensions \cite{Romero98g,Romero98d,Romero99b}:
\begin{equation}
C_{\vec{r}}^{[D]} = \langle \hat{C}_{\vec{r}}^{[D]} \rangle = \frac 1 {2g} \sum_{\vec{n}} \langle a_{\vec{n}}^{\dagger} a_{\vec{n}} (b_{\vec{n}+\vec{r}}^{\dagger} + b_{\vec{n}+\vec{r}} ) \rangle  ~,
\label{eq:corrfunc}
\end{equation}
normalized such that $\sum_{\vec{r}} C_{\vec{r}}^{[D]} = 1$.
This function can be viewed as measuring the {\it shape} of the polaron lattice distortion around the instantaneous position of the electron.

Using Rayleigh-Schr\"{o}dinger perturbation theory in the weak-coupling regime as in the preceeding sections one finds that
\cite{Romero98g,Romero98d}
\begin{eqnarray}
{C}_{\vec{r}}^{[D]} &=& \hbar \omega \int_0^{\infty} \!\! dt ~ e^{-\hbar\omega t} \prod_{i=1}^D e^{- 2J_i t} I_{r_i} (2J_i t)  ~.
\label{eq:c3d}
\end{eqnarray}
Note that setting any one $J_i$ to zero or summing $ {C}_{\vec{r}}^{[D]} $ over one $r_i$ recovers ${C}_{\vec{r}}^{[D-1]} $.
This property implies that the effect of ``turning on'' transverse dimensions is simply to spread electron-phonon correlation strength transversely.

Characterizing this multi-dimensional correlation function in terms of a width measure involves a variance tensor,

\begin{equation}
\left\{ {\bf \sigma^2} \right\}_{ij} = \sum_{\vec{r}} r_i r_j {C}_{\vec{r}}^{[D]} = \delta_{ij} \sigma_{ii}^2 ~,
\label{eq:vardef}
\end{equation}
where
\begin{eqnarray}
\sigma_{ii}^2 &=& \hbar \omega \int_0^{\infty} \!\! dt ~ e^{- \hbar \omega t} ~ \sum_{r_i} ~ r_i^2 \, e^{- 2J_i t} I_{r_i} (2J_i t) \\
&=& \frac {2J_i} {\hbar \omega} = \frac \hbar {2 m_{ii}^0 \omega} \frac 1 { l_i^2} ~.
\label{eq:varwcpt}
\end{eqnarray}
in which $m_{ii}^0$ is the free electron effective mass and $l_i$ is the lattice constant in the $i$ direction.
Thus, along each of the primitive crystallographic axes, the real-space variance is simply proportional to the electron transfer integral along that axis, and in a general direction is just the appropriate mixture determined by rotation.
In absolute units (unrationalized by the lattice constants) the real-space variance is the same as that of the zero-point motion of a harmonic oscillator  characterized by the lattice frequency $\omega$, but with the lattice mass replaced by the free electron mass measured along the appropriate direction.

Utilizing the notion of a polaron half-width defined in terms of the correlation variance
\begin{equation}
R_i = \frac {l_i} 2 \sqrt{ \sigma_{ii}^2 } ~,
\label{eq:width}
\end{equation}
we can associate with the polaron characteristic ellipsoidal volumes $V[D]$
\begin{eqnarray}
V [1] &\sim& 2 R_x ~\sim~ l_x \left( \frac {2J_x} {\hbar\omega} \right)^{1/2}
\nonumber \\
&\sim& \left( \frac \hbar { 2 \omega } \right)^{1/2} \left( \frac 1 {m_{ii}^0} \right)^{1/2} \\
V [2] &\sim& \pi R_x R_y ~\sim~ l_x l_y \frac {\pi} 4 \left( \frac {2J_x} {\hbar\omega} \frac {2J_y} {\hbar\omega} \right)^{1/2}
\nonumber \\
& \sim & \frac {\pi} 4 \left( \frac \hbar { 2 \omega } \right) \left( det ~ {\bf M_0^{-1}} \right)^{1/2} ~. \\
V [3] &\sim& \frac {4 \pi} 3 R_x R_y R_z ~\sim~ l_x l_y l_z \frac {\pi} 6 \left( \frac {2J_x} {\hbar\omega} \frac {2J_y} {\hbar\omega} \frac {2J_z} {\hbar\omega} \right)^{1/2}
\nonumber \\
& \sim & \frac {\pi} 6 \left( \frac \hbar { 2 \omega } \right)^{3/2} \left( det ~ {\bf M_0^{-1}} \right)^{1/2} ~.
\label{eq:volume}
\end{eqnarray}
This characteristic volume thus increases with the intensity of tunneling ($V[D] \propto |\vec{J} /\hbar\omega | ^{D/2}$), and is largest in the isotropic case and decreases with increasing anisotropy.
In the isotropic case we may regard $R = R_i$ as the polaron radius.

Contrary to much prevailing opinion, these results show that in the weak-coupling regime:
i)~there are no significant qualitative or quantitative differences between 1D, 2D, and 3D polaron radii,
ii)~the polaron radius in 2D and 3D is {\it not} infinite, and
iii)~the polaron radius does not scale as $J/g^2 \hbar \omega$ in {\it any} dimension as commonly expected, but as $\sqrt{J/ \hbar \omega}$ in {\it every} dimension \cite{Romero98g,Romero98d,Romero99b}.

\section{Effective Mass}
\label{sec:mass}

For the circumstances we address in this paper, the reciprocal effective mass tensor is diagonal, with elements given by
\begin{equation}
\left\{ {\bf M^{-1}} \right\}_{ij} = \left. \hbar^{-2} \frac {\partial^2 E( \vec{\kappa} )} {\partial \kappa_i \partial \kappa_j } \right|_{\vec{\kappa} = 0} = \delta_{ij} \frac 1 {m_{ii}} ~.
\label{eq:masstensor}
\end{equation}
From this, it is easily shown that the reciprocal effective mass in any direction through second order of weak-coupling perturbation theory is given by
\begin{equation}
\frac {m_{ii}^0} {m_{ii}^*} = 1 - g^2 \hbar^2 \omega^2 \!\! \int_0^{\infty} \!\! \!\! \!\! dt ~ t ~ e^{- \hbar \omega t} \prod_{i=1}^D e^{- 2J_i t} I_0 (2J_i t) ,
\label{eq:wcptmass}
\end{equation}
where $m_{ii}^*$ and $m_{ii}^0$ are respectively the polaron and free electron effective masses in the $i$ direction..

\begin{figure}[htb]
\begin{center}
\leavevmode
\epsfxsize = 3.2in
\epsffile{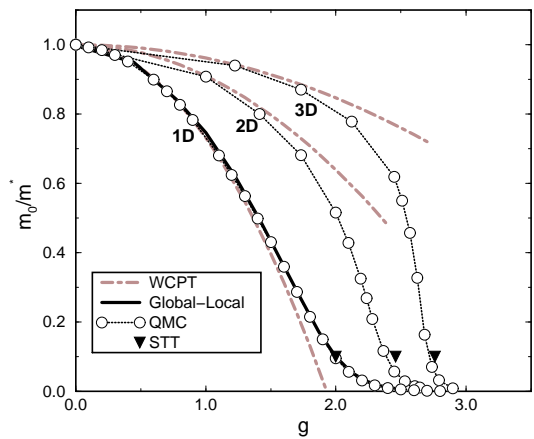}
\end{center}
\caption{
The reciprocal effective mass ratio $m_{ii}^0 / m_{ii}^*$ for the isotropic case.
Solid curve:
Global-Local mass for $D=1$.
Chain-dashed curves:
WCPT masses for $D=1$, $2$, and $3$.
Scatter-plot:
Quantum Monte Carlo data for $D=1$, $2$, and $3$; data kindly provided by P. E. Kornilovitch \protect \cite{Kornilovitch98a,Kornilovitch99} \protect .
Triangles:  Estimated self-trapping points $g_{ST}$ as discussed in Section \ref{sec:finite}.
}
\label{fig:mass123d}
\end{figure}

Figure~\ref{fig:mass123d} shows the dependence of the {\it isotropic} polaron mass on dimensionality according to WCPT and quantum Monte Carlo simulation.
Although this is a comparison between isotropic cases of $J/\hbar\omega = 1$ only, the excellent agreement between WCPT and quantum Monte Carlo out to $g \sim \sqrt{D}$ suggests that the WCPT mass may be similarly accurate for $\lambda < 1/2$ as defined in (\ref{eq:lambdadef}) at general $\vec{J}$.

\begin{figure}[htb]
\begin{center}
\leavevmode
\epsfxsize = 3.2in
\epsffile{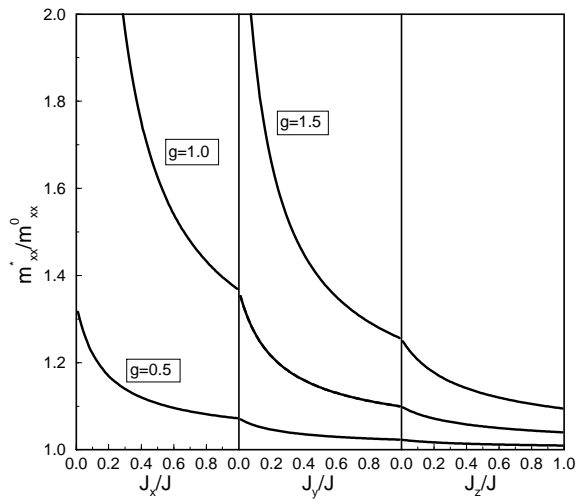}
\end{center}
\caption{
The effective mass ratio $m_{xx}^* / m_{xx}^0$ according to (\ref{eq:wcptmass}), showing the effects of changing anisotropy and dimensionality.
Left Panel:
1D mass ($J_y = J_z = 0$) at fixed $g$, increasing $J_x /\hbar\omega$ from $0$ to $1$.
Center Panel:
2D mass ($J_x / \hbar \omega, ~J_z / \hbar \omega = 0$) at fixed $g$, increasing $J_y /\hbar\omega$ from $0$ to $1$.
Right Panel:
3D mass ($J_x / \hbar \omega = J_y / \hbar \omega = 1$) at fixed $g$, increasing $J_z /\hbar\omega$ from $0$ to $1$.
}
\label{fig:emjxjyjz}
\end{figure}

The weak-coupling result (\ref{eq:wcptmass}) shows that although anisotropy has definite effects on the value of the effective mass, the effect of anisotropy appears only in the value of a scalar multipler of the free electron mass; that is,
although anisotropy of the free electron mass (inequalities among $J_x$, $J_y$, and $J_z$) is manifested in real-space anisotropies in electron-phonon correlation (i.e., in distortions of the {\it shape} of the polaron as discussed in the previous section),
 the mass renormalization associated with such distortions of polaron shape is {\it isotropic}.
Interestingly, this implies that increasing $J_y$ or $J_z$ at fixed $J_x$ (for example) results in a {\it decrease} in $m_{xx}^*$, translating into an associated increase in mobility in the $x$ direction.
This influence of transverse directions on $m_{xx}^*$ is illustrated in Figure~\ref{fig:emjxjyjz}.
In the center and right panels of Figure~\ref{fig:emjxjyjz}, $J_x / \hbar \omega$ is held fixed at unity, yet the effective mass in the $x$ direction continues to decrease as tunneling into transverse dimensions is turned on.

These effects can be understood in terms of the transverse spreading of electron-phonon correlation strength as discussed in the last section.
As a fixed correlation strength is spread over an increasing number of sites (characteristic volume of the polaron increases as discussed in the previous section), the average lattice deformation per participating site decreases.
Consequently, mean square measures of lattice deformation decrease and exhibit changes that suggest a diminishing effectiveness of electron-phonon interactions in producing typical polaronic effects.
The polaronic mass enhancement bears such a mean square dependence on the lattice deformation and like other such measures (e.g., the number of phonons in the phonon cloud as discussed in Appendix A), decreases with increasing dimensionality and decreasing anisotropy.

The corresponding polaron effective mass resulting from strong-coupling perturbation theory through second order is given by
\begin{equation}
\frac {m_{ii}^0} {m_{ii}^*} = e^{-g^2} + e^{-2g^2} f(g^2) ( 3J_i + {\cal{J}} )/\hbar\omega ~.
\label{eq:scptmass}
\end{equation}
This result is isotropic at first order simply by virtue of being independent of all $J_i$ at that order, but the second-order correction is anisotropic because the r.h.s. of (\ref{eq:scptmass}) bears an explicit, unbalanced sensitivity to the direction along which the effective mass component is being measured.

Unfortunately, this strong-coupling result is not very helpful; it disagrees substantially with more reliable results \cite{Jeckelmann98a,Kornilovitch98a,Kornilovitch99,Romero98e} except at small $J/\hbar\omega$.
We take this as an indication that dominating (perhaps non-exponential) contributions have yet to be extracted from higher orders of SCPT.
For such reasons we cannot estimate the location of the self-trapping transition from any crossing of (\ref{eq:wcptmass}) and (\ref{eq:scptmass}).
Instead, we have included in Figure~\ref{fig:mass123d} several symbols to indicate the values of $g_{ST}$ as given by (\ref{eq:gstd}); these several values are mutually consistent in locating essentially the same feature of the effective mass in every dimension, and coincides very well with the effective mass feature we have previously identified with the self-trapping transition (see Ref. \cite{Romero98e}).

\section{On Dimensionality and Adiabaticity}

As noted in the introduction, the results that have long characterized commonly-held expectations for the dimensionality dependence of polaron structure are due to behavior ascertainable in the adiabatic approximation
\cite{Rashba57a,Rashba57b,Holstein59a,Derrick62,Emin73,Sumi73,Emin76,Toyozawa80a,Schuttler86,Ueta86,Kabanov93,Silinsh94,Song96,Holstein81,Holstein88a,Holstein88b}.

In 2D and 3D, the minimum energy states in the adiabatic approximation are found to be ``free'' states throughout the weak-coupling regime up to a discrete coupling threshold beyond which ``self-trapped'' states have the minimum energy.
This abrupt transition phenomenon is what is meant by the term ``self-trapping transition'' in the adiabatic approximation.
Accordingly, there is no occasion to distinguish large polarons from small polarons in 2D and 3D since the ``free'' states below the transition are of infinite radius and {\it distinct} from large polarons, and the ``self-trapped'' states above the transition are always interpretable as small polarons.
This set of circumstances in 2D and 3D is reflected in the catch phrase ``all polarons are small'', since in this view large polarons in the adiabatic sense are never characteristic of the polaron ground state in bulk materials.

In 1D, on the other hand, ``free'' states are unstable in the adiabatic approximation; instead, finite-radius (i.e. ``self-trapped'') states are found at all finite coupling strengths, leading to the commonly encountered view that there is no self-trapping transition in 1D.
That polaron states in 1D might be distinguishable as large or small is inconsequential in this view, as is the notion of a resolvable transition between distinct large and small polaron structures.

The results of this paper differ strongly from the conventional adiabatic picture in multiple respects:

i) The quasiparticles implicit in the weak-coupling states of {\it every} dimension are not weakly-scattered ``free'' electrons, but dressed electrons having finite radii generally greater than a lattice constant.

ii) Although these weak-coupling quasiparticles can be sensibly characterized as large polarons, in {\it no} dimension do these weak-coupling states coincide with the large polaron states familiar from the adiabatic approximation in 1D.

iii) The finite radii characterizing the weak-coupling quasiparticles in {\it every} dimension saturate to finite values with vanishing electron-phonon coupling, unlike the large polaron radii in the adiabatic approximation that in 1D diverge with vanishing coupling and in 2D and 3D are infinite already at finite coupling.

iv) The self-trapping transition exists in {\it every} dimension, including 1D.

v) The self-trapping transition is associated with the change from large polaron structure to small polaron structure in {\it every} dimension, including 2D and 3D, and {\it not} with a change from infinite to finite radii.

vi) Dependences of polaron properties on parameters are smooth through the self-trapping transition in {\it every} dimension, unlike the abrupt changes often found in the adiabatic approximation in 2D and 3D.

Our results are quantitatively supported by independent high-quality methods (including variational methods \cite{Romero98g,Romero98c,Romero98e,Romero98a,Brown97b,Trugman99}, cluster diagonalization \cite{Capone97,Wellein97a,deMello97,Alexandrov94a}, density matrix renormalization group \cite{Jeckelmann98a}, and quantum Monte Carlo \cite{DeRaedt83,DeRaedt84,Lagendijk85,Kornilovitch98a,Kornilovitch99,Alexandrov98a}).
Moreover, elaborations of adiabatic theory incorporating {\it non-adiabatic} corrections \cite{Kabanov93,Alexandrov94a,Kabanov94} support our overall conclusion that the adiabatic approximation as it is widely regarded fails to embrace non-adiabatic characteristics that are essential to the proper description of polaron states in the weak coupling regime, and therefore fails as well to properly describe the self-trapping transition itself \cite{Romero98f}.

With so many results at variance with the adiabatic approximation, it is well to ask in what respects, if any, are our results {\it consistent} with the adiabatic approximation and whether some sense can be made of the pervasive discrepancies.
Indeed, several consistencies can be found that are illuminating.

We first note that the dependence of $\lambda_c$ on dimensionality and anisotropy exhibits a generic square-root singularity at the boundary between 1D and any higher dimensional case, while at the boundary between higher-dimensional cases this dependence is generically smooth.
For essentially the same underlying reasons, $\lambda_c$ is constant throughout 1D, but varies with detail of tunneling in higher dimensions.
These distinctions are at least suggestive of the sharp contrasts between 1D and higher dimensional cases in the adiabatic approximation.

Secondly, we note that the weak-coupling polaron radius $R$ as here derived diverges in any dimension in the adiabatic limit.
Further considering the WCPT validity test in Appendix A, there is reason to speculate that this weak-coupling radius might continue to be a reasonably valid construct in 2D and 3D up to the vicinity of the self-trapping transition.
Such a possibility might be consistent with the finding of strictly infinite-radius states on the weak-coupling side of the transition in 2D and 3D in the adiabatic approximation, while the possible breakdown of the weak-coupling radius construct below the transition in 1D might be consistent with existence of finite-width states in the 1D adiabatic approximation.

\section{Conclusion}
\label{sec:concl}

In this paper we have analyzed the dependence of numerous polaron properties on the effective real-space dimensionality and anisotropy as determined by the electronic tunnelling matrix elements; these properties include the polaron ground state energy, polaron shape, size, and volume, the number of phonons in the phonon cloud and the polaron effective mass.
In pursuing these analyses we have made extensive use of weak- and strong-coupling perturbation theories supported by selected comparisons with non-perturbative methods.
Through the use of a scaling argument combining weak- and strong-coupling perturbation theory in the adiabatic strong-coupling regime, we have been able to infer the probable location of the self-trapping critical point in the adiabatic limit in any dimension and for any degree of anisotropy, and by combining information from multiple sources we have been able to extend this estimate from the adiabatic limit to finite adiabaticity.

Central among our findings is the over-arching qualitative conclusion that polarons in any dimension and any degree of anisotropy are similar in most respects.
In particular, polarons on the weak-coupling side of the self-trapping transition share a structure that is essentially identical in every dimension.
This weak-coupling structure is consistent with the notion of the weak-coupling polaron as a finite-radius quasiparticle, but is {\it inconsistent} both with the notion of a weakly-scattered free electron (adiabatic approximation in 2D and 3D) and with the historical notion of the large polaron (adiabatic approximation in 1D).
The strong-coupling structure is consistent with traditional notions of small polarons, including strong-coupling perturbation theory and the adiabatic approximation.

Since the essential character of the weak-coupling states and strong-coupling states is only inessentially affected by dimensionality and anisotropy, the notion of the self-trapping transition separating the weak- and strong-coupling states is similarly not altered in any essential way by changes in dimensionality or anisotropy.
Necessarily, one is led to view self-trapping as the more-or-less rapid transition, occurring in {\it every} dimension, between characteristic weak- and strong-coupling states, both of which are characterized by finite radii.

If we may transcend the jargon that historically has had a tendency to polarize the conventional wisdom, it is fairly concluded that {\it not all polarons are small}, even in bulk materials, and that in every dimension and for every degree of anisotropy the self-trapping transition is a smooth, albeit rapid crossover between large and small polaron character.

\section*{Acknowledgement}

The authors gratefully acknowledge P. Kornilovitch for providing the quantum Monte Carlo data used in Figures~\ref{fig:grnd123d} and \ref{fig:mass123d}.
This work was supported in part by the U.S. Department of Energy under Grant No. DE-FG03-86ER13606.

\appendix

\section{Breakdown of Weak-Coupling Perturbation Theory}
\label{sex:append}

The weak-coupling perturbation theory considered in this paper is based on an expansion in states containing limited numbers of phonon quanta.
The zeroth order properties are based upon states containing zero phonons, and second order properties upon states containing one phonon.
The first neglected order of WCPT is the fourth order, built upon states containing no more than two phonons.
A test of internal consistency of WCPT at particular parameters, therefore, is to compute the expected number of phonons to the retained order of perturbation theory, and compare this number to the maximum number of phonons present at that order.
For second order WCPT as used in this paper, this number of phonons should be less than unity.

The required computation is contained in
\begin{equation}
n_{ph} = \frac 1 N \sum_{\vec{q}} \frac {g^2 \hbar \omega}
{\{ E_{WC}^{(0)} (0) - [ E_{WC}^{(0)} (- \vec{q} ) + \hbar \omega ] \} ^2} ~,
\end{equation}
where $E_{WC}^{(0)} ( \vec{\kappa} )$ is defined in (\ref{eq:wcpt2}).
When {\it each} $J_i$ is {\it large} relative to $\hbar \omega$, one finds that
\begin{eqnarray}
n_{ph} &\sim& \frac 1 4 g^2 \left( \frac {J_x} {\hbar \omega} \right)^{-1/2} ~ ~~~~~ ~~~~~ in~1D ~,\\
&\sim& \frac 1 \pi g^2 \left( \frac {J_x J_y} {\hbar^2 \omega^2} \right)^{-1/2} ~~~ ~~~~~ in~2D ~, \\
&\sim& \frac 1 \pi g^2 \left( \frac {J_x J_y J_z} {\hbar^3 \omega^3} \right)^{- 1/2} ~~~~~ in~3D ~.
\end{eqnarray}
These expressions can be consolidated into the single approximate relation
\begin{equation}
n_{ph} \propto g^2 \frac {\Omega [D]} {V [D]} ~,
\end{equation}
where $\Omega [D]$ is the primitive cell volume and $V[D]$ the characteristic volume of the polaron in $D$ dimensions.
The dimension-dependent constant of proportionality is near 1/2 in all cases.
This simple relation, here proven only for the adiabatic weak-coupling regime (broad polarons), demonstrates the very direct but inverse relation between the number of phonons in the phonon cloud and the volume occupied by it.

In the isotropic case and in terms of the composite parameter $\lambda$, the condition that expected phonon numbers should be less than unity results in the conditions ($J \gg \hbar \omega$)

\begin{eqnarray}
\lambda &<& 2 ~ \left( \frac J {\hbar \omega} \right) ^{-1/2} ~~~~~ in~1D ~,\\
&<& \frac \pi 4 ~~~~~ ~~~~~ ~~~~~ ~~~~~ in~2D ~, \\
&<& \frac \pi 6~ \left( \frac J {\hbar \omega} \right) ^{1/2} ~ ~~~~~ in~3D ~.
\end{eqnarray}

Recalling that the self-trapping transition is expected to occur at $\lambda$ of order unity, it would appear that WCPT through second order is consistent with the condition $n_{ph} < 1$ up to the transition in 2D and beyond the transition in 3D.
It is the 1D case that appears to be on the weakest footing in the adiabatic strong-coupling regime; however, it is the 1D case that has been most exhaustively studied by non-perturbative means and found to be widely consistent with second-order WCPT.

\bibliography{/home/bassi/dwb/Tex/Bibliography/theory,/home/bassi/dwb/Tex/Bibliography/books,/home/bassi/dwb/Tex/Bibliography/experiment,/home/bassi/dwb/Tex/Bibliography/temporary}

\end{document}